# Some exact solutions in Møller gravity

**Edward Rakhmetov**
*Skobeltsyn Institute of Nuclear Physics, M.V. Lomonosov Moscow State University*
*1(2), Leninskie gory, GSP-1, Moscow 119991, Russian Federation*
*E-mail: dsparel@mail.ru*

**Sergey Keyzerov**
*Skobeltsyn Institute of Nuclear Physics, M.V. Lomonosov Moscow State University*
*1(2), Leninskie gory, GSP-1, Moscow 119991, Russian Federation*
*E-mail: errar@www-hep.sinp.msu.ru*

We present a short introduction to the Møller gravity theory and study the Schwarzschild solution and self-consistent spontaneous compactification solutions in Kaluza-Klein theories, which can be obtained in the framework of this interesting generalization of General Relativity.

## 1. Introduction

In this paper we describe an interesting generalization of General Relativity suggested by C. Møller [1]. Save for this little introduction and a brief review of the Møller gravity, we are going to discuss exact solutions, which can be obtained in this theory. Namely, the Schwarzschild solution and self-consistent spontaneous compactification solutions in Kaluza-Klein theories.

We will look for self-consistent solutions of the form $M_4 \times S^n$, where $M_4$ is the 4-dimensional Minkowski space and $S^n$ is an n-dimensional sphere. Such solutions are useful to obtain, after a dimensional reduction, effective 4-dimentional theories with CP-violation, which might be very interesting in particle and high energy physics [2,3]. But unfortunately, such solutions cannot be found in General Relativity (except the case n=1), so that we must consider other theories.

The next motivation is an attempt to explain the modern astrophysics data (such as the rotation curves of stars in spiral galaxies) by means of a modification of gravity on large scales without involving the "Dark Matter" concept [4,5].

And last, but not least: we look for a theory with a weak Lorentz symmetry violation. In the Møller gravity theory, we have more restrictions on the frame vectors, than in General Relativity, because of an additional antisymmetric part of the equations of motion. Thus, we have a hope to find a Lorentz symmetry violation in some cases, when the frame vectors give us a preferred direction in space-time because of these restrictions. In other words, we can obtain a weak Lorentz symmetry violation, because the action of the theory is not invariant under the rotations of the frame vectors in some cases. But we are not going to discuss such cases in this paper.

## 2. Møller gravity theory

It was C. Møller, who first put forward this theory in 1978 [1]. The Møller gravity theory is a metric theory, in which the metric tensor

$$g_{\mu\nu} \equiv \sum_{\alpha=0}^{3} \delta(\alpha\alpha) g(\alpha)_\mu g(\alpha)_\nu \equiv g(\alpha)_\mu g(\alpha)_\nu, \quad \delta(\alpha\beta) \equiv diag\{-1,1,1,1\} \tag{2.1}$$

is constructed from an orthonormal tetrad (vielbein):

$$g(\alpha)_\mu g(\beta)^\mu \equiv \delta(\alpha\beta) \tag{2.2}$$

Here indices in the brackets are frame (vielbein) indices, which run from 1 to 4, and the summation over the repeated indices is implied. The stress tensor for this vector fields (denoted by $f(\alpha)_{\mu\nu}$) is, as usually,

$$f(\alpha)_{\mu\nu} \equiv \partial_{[\mu} g(\alpha)_{\nu]}, \tag{2.3}$$

where the antisymmetrization over the indices in the square brackets is implied. The coordinate indices can be turned into vielbein indices as it is presented below:

$$C_{\mu\nu}(\alpha) g(\alpha)_\lambda \equiv C_{\mu\nu\lambda} \tag{2.4}$$

We can obtain the Ricci tensor from the stress tensor:

$$R(\alpha\beta) = \frac{1}{2}\partial(\alpha) f(\mu\mu\beta) + \frac{1}{2}\partial(\beta) f(\mu\mu\alpha) + \frac{1}{2}\partial(\mu) f(\beta\alpha\mu)$$
$$+ \frac{1}{2}\partial(\mu) f(\alpha\beta\mu) - \frac{1}{2} f(\alpha\beta\mu) f(\nu\nu\mu) - \frac{1}{2} f(\beta\alpha\mu) f(\nu\nu\mu) + \tag{2.5}$$
$$- \frac{1}{2} f(\mu\nu\alpha) f(\nu\mu\beta) - \frac{1}{2} f(\mu\nu\alpha) f(\mu\nu\beta) + \frac{1}{4} f(\alpha\mu\nu) f(\beta\mu\nu)$$

where $\partial(\alpha) \equiv g(\alpha)^\mu \partial_\mu$. \hfill (2.6)

Contacting indices in different ways, we can build from the stress tensor three different scalars:

$$L_1 \equiv f_{\alpha\beta\gamma} f^{\alpha\beta\gamma} \quad L_2 \equiv f_{\alpha\beta\gamma} f^{\beta\alpha\gamma} \quad L_3 \equiv f^\alpha{}_{\alpha\gamma} f_\beta{}^{\beta\gamma} \tag{2.7}$$

Then, in the general case, the simplest action quadratic in the partial derivatives is:

$$S = \int_X \left(k_0 + k_1 L_1 + k_2 L_2 + k_3 L_3\right) \sqrt{g}\, dx, \tag{2.8}$$

where $k_0$, $k_1$, $k_2$, $k_3$ are arbitrary dimensional constants.

Using (2.5) we can obtain a more usual expression for the Einstein-Møller action (up to a complete divergence):

$$S = \int_X \left(k_0 + k_3 R + k_1' L_1 + k_2' L_2\right) \sqrt{g}\, dx \tag{2.9}$$

We could not write the action with the terms $L_1$ and $L_2$, if we used just the metric tensor without the vielbein formalism. Thus, we have a generalization of metric gravity.

As we can see, the Møller gravity theory coincides with General Relativity, if $k_1'$, $k_2'$ are equal to 0.

Denoting the variation of the action with respect to $g(\mu)_\alpha$ by $X(\mu)^\alpha \equiv \frac{1}{\sqrt{g}} \frac{\delta S}{\delta g(\mu)_\alpha}$,

we can write the symmetric and the antisymmetric parts of the equations of motion

$X(\mu)^\alpha \equiv \frac{1}{\sqrt{g}} \frac{\delta S}{\delta g(\mu)_\alpha} = 0$ separately. The symmetric part gives:

$$X^{(\alpha\beta)} = -4k_3\left(R^{\alpha\beta} - \frac{1}{2}g^{\alpha\beta}(R+\Lambda)\right) + 2(2k_1' + k_2')\nabla_\mu f^{(\alpha\beta)\mu} - (2k_1' - k_2')f^{[\mu\nu]\alpha}f_{[\mu\nu]}{}^\beta +$$
$$+ (2k_1' + k_2')f^{\mu\nu(\alpha}f^{\beta)}{}_{\mu\nu} - 2k_2' f^\alpha{}_{\mu\nu}f^{\beta\mu\nu} + 2g^{\alpha\beta}\left(k_1' f^{\mu\nu\lambda}f_{\mu\nu\lambda} + k_2' f^{\mu\nu\lambda}f_{\nu\mu\lambda}\right) = 0 \quad (2.11)$$

The antisymmetric part of the equations of motion is

$$X^{[\alpha\beta]} = 2(2k_1' - k_2')\nabla_\mu f^{[\alpha\beta]\mu} - 4k_2'\nabla_\mu f^{\mu\alpha\beta} - (2k_1' - 3k_2')f^{\mu\nu[\alpha}f^{\beta]}{}_{\mu\nu} = 0 \quad (2.12)$$

If $k_1'$, $k_2'$ are equal to 0, the antisymmetric part vanishes, and the symmetric part gives us General Relativity. As we can see, in the Møller gravity theory we have more restrictions on the frame vectors, than in General Relativity, because of the additional antisymmetric part of the equations of motion.

## 3. Schwarzschild solution in Møller gravity

In this section small Latin indices run from 1 to 3. Let us write the well-known Schwarzschild metric in the following way:

$$ds^2 = -e^{2\gamma(r)}dt^2 + e^{2\alpha(r)}\left\{dr^2 + r^2\left(d\varphi^2 + \sin^2\varphi d\chi^2\right)\right\} \quad (3.1)$$

Thus, the ansatz for the metric tensor is:

$$g_{\alpha\beta} = \begin{pmatrix} -e^{2\gamma} & 0 \\ 0 & e^{2\alpha}\bar{g}_{pq} \end{pmatrix} \quad (3.2)$$

The ansatz for the vielbein looks like:

$$g(0)_0 = e^\gamma \qquad g(a)_q = e^\alpha \bar{g}(a)_q \qquad g(0)_q = 0 \qquad g(a)_0 = 0 \quad (3.3)$$
$$g(0)^0 = e^{-\gamma} \qquad g(a)^q = e^{-\alpha}\bar{g}(a)^q$$

where $g(a)^q$ is an orthonormal frame in 3-dimensional flat space and the 3-vectors $g(a)^q$ are defined as follows

$$\bar{g}(1)_q = \begin{pmatrix} \cos\varphi \\ -r\sin\varphi \\ 0 \end{pmatrix} \qquad \bar{g}(2)_q = \begin{pmatrix} -\sin\varphi\cos\chi \\ -r\cos\varphi\cos\chi \\ r\sin\varphi\sin\chi \end{pmatrix} \qquad \bar{g}(3)_q = \begin{pmatrix} -\sin\varphi\sin\chi \\ -r\cos\varphi\sin\chi \\ -r\sin\varphi\cos\chi \end{pmatrix} \quad (3.4)$$

After some boring calculations we obtain three equations for two functions parameterizing the metric and the vielbein:

$$2\alpha_{,rr} + \alpha_{,r}^2 + 4r^{-1}\alpha_{,r} = \frac{\Lambda}{2}e^{2\alpha} + \kappa\left(2\alpha_{,r}^2 - 2\gamma_{,rr} - \gamma_{,r}^2 - 2\alpha_{,r}\gamma_{,r} - 4r^{-1}\gamma_{,r}\right) \quad (3.5)$$

$$\alpha_{,r}^2 + 2r^{-1}\alpha_{,r} + 2\gamma_{,r}\alpha_{,r} + 2r^{-1}\gamma_{,r} = \frac{\Lambda}{2}e^{2\alpha} - \kappa\left(2\alpha_{,r}^2 + \gamma_{,r}^2 + 4r^{-1}\alpha_{,r}\right) \quad (3.6)$$

$$\alpha_{,rr} + r^{-1}\alpha_{,r} + r^{-1}\gamma_{,r} + \gamma_{,rr} + \gamma_{,r}^2 = \frac{1}{2}e^{2\alpha}\Lambda + \kappa\left(-2\alpha_{,rr} + \gamma_{,r}^2 - 2\alpha_{,r}\gamma_{,r} - 2r^{-1}\alpha_{,r}\right) \quad (3.7)$$

where $\kappa \equiv \dfrac{2k_1' + k_2'}{2k_3}$ \quad (3.8)

Thus, the system is either overdetermined, or these equations are not independent. It can be shown that the system has solutions only if $\kappa = 0$. But in this case we have the same system as in General Relativity! In other words, the Møller theory has the same Schwarzschild solution as General Relativity. Then a peculiar point is that this solution appears not only in the case, where the constants $k_1$ and $k_2$ are small, as it was shown by Møller [1], but also in the case of arbitrary constants too, when the relation $2k'_1 + k'_2 = 0$ is valid. If this relation is not valid, there are no spherically symmetric Schwarzschild-like solutions in the Møller gravity theory or, at any rate, we cannot find a way neither to discribe, nor to obtain them.

## 4. Self-consistent spontaneous compactification solutions of the form $M_4{}^x S^3$

In this section large Latin indices take the values from 0 to 7, with the exception of 4, small Latin indices run from 0 to 3 and small Greek indices run from 5 to 7. As it was mentioned above the space-time manifold is $M_4{}^x S^3$. The ansatz for the vielbein looks as follows:

$$g(\alpha)_\mu(x^A) = h(\alpha)_\mu(x^\lambda) \quad \text{depend only on the 4-dimensional coordinates} \tag{4.1}$$

$$g(\alpha)_q = 0 \tag{4.2}$$

$$g(a)_\mu = 0 \tag{4.3}$$

$$g(a)_q(x^A) = r\bar{g}(a-4)_q(x^l) \quad \text{depend only on the coordinates on } S^3 \tag{4.4}$$

Here $h(\alpha)_\mu$ is a vielbein in Minkowski space $M_4$ and $\bar{g}(n)_q$ is a vielbein on 3-dimensional sphere $S^3$. Then there are two nontrivial equations of motion:

$$R[h](\alpha\beta) - \frac{1}{2}\left(R[h] + 6r^{-2} - \Lambda\right)\delta(\alpha\beta) = \frac{1}{4k_3}Y[h](\alpha\beta) \tag{4.6}$$

$$2r^{-2}\delta(pq) - \frac{1}{2}\left(R[h] + 6r^{-2} - \Lambda\right)\delta(pq) = 4\kappa r^{-2}\delta(pq) - 6\frac{k'_2}{k_3}r^{-2}\delta(pq) \tag{4.7}$$

where $Y^{\alpha\beta} \equiv X^{(\alpha\beta)} + 4k_3\left(R^{\alpha\beta} - \frac{1}{2}g^{\alpha\beta}(R + \Lambda)\right)$ (4.8)

For Minkowski space $R[h](\alpha\beta) = 0$, so that we get from equations (4.6)-(4.7) compatibility conditions for the constants $k'_1, k'_2$ and the compactification radius r

$$k'_2 = -\frac{1}{3}k_3 \qquad k'_1 = -\frac{1}{6}k_3 \qquad r^{-2} = \frac{1}{6}\Lambda \tag{4.9}$$

If we take such values of the constants, we get a solution for the metric tensor corresponding to the desired direct product structure $M_4{}^x S^3$. That is to say, with such values of the constants the four-dimensional dynamics allows solutions with flat Minkowski space and the other dimensions compactified into 3-dimensional sphere.

## 5. Conclusions

In the Møller gravity the Schwarzschild solution appears not only in the case, where the additional constants of the Einstein-Møller action are small, as it was shown by Møller [1], but also in the case of arbitrary constants too, when certain relations between these constants are valid. If these relations are not valid, there are no spherically symmetric Schwarzschild-like solutions in the Møller gravity theory or we need some other way to discribe them.

The four-dimensional dynamics allows solutions with flat Minkowski space for a wide range of the parameters of the theory, when the extra dimensions are spontaneously compactified into 3-dimensional sphere. This is very interesting, because in General Relativity solutions of this type are impossible without additional vector or tensor fields [6].

The action of the theory is not invariant under the rotations of the frame vectors. This can result in a theory with a violation of the Lorentz symmetry, when there are certain relations between the parameters of the theory, but in this paper we do not consider such cases.

**Acknowledgements**

The authors are grateful to I.P. Volobuev for valuable discussions. The work was supported by FASI state contract 02.740.11.0244 and by grant of Russian Ministry of Education and Science NS-4142.2010.2.